\documentclass[%
 reprint,
 amsmath,amssymb,
 aps,
]{revtex4-2}

\usepackage{graphicx}
\usepackage{dcolumn}
\usepackage{bm}
\usepackage{physics}

\usepackage[T1]{fontenc}
\usepackage[usenames,dvipsnames]{xcolor}
\usepackage{hyperref}
\usepackage[version=3]{mhchem}

\usepackage{threeparttable}
\usepackage{longtable}

\newcommand{\jb}{\bar{j}}
\newcommand{\kb}{\bar{k}}

\newcommand{\bb}{\bar{b}}
\newcommand{\cb}{\bar{c}}
\newcommand{\rmh}{\textrm{h}}

\begin{document}


\title{Exploring electron affinities, LUMO energies, and band gaps with electron-pair theories}

\author{Marta Gałyńska}
 \affiliation{Institute of Physics, Faculty of Physics, Astronomy, and Informatics, Nicolaus Copernicus University in Toruń, Grudziądzka 5, 87-100 Toruń, Poland}
\author{Pawe\l{} Tecmer}%
 \affiliation{Institute of Physics, Faculty of Physics, Astronomy, and Informatics, Nicolaus Copernicus University in Toruń, Grudziądzka 5, 87-100 Toruń, Poland}
\author{Katharina Boguslawski}%
 \email{k.boguslawski@fizyka.umk.pl}
 \affiliation{Institute of Physics, Faculty of Physics, Astronomy, and Informatics, Nicolaus Copernicus University in Toruń, Grudziądzka 5, 87-100 Toruń, Poland}


\date{\today}

\begin{abstract}
We introduce the electron attachment equation-of-motion pair coupled cluster doubles (EA-EOM-pCCD) ansatz, which allows us to inexpensively compute electron affinities, energies of unoccupied orbitals, and electron attachment spectra.
We assess the accuracy of EA-EOM-pCCD for a representative data set of organic molecules for which experimental data is available, as well as the electron attachment process in uranyl dichloride.
EA-EOM-pCCD provides more reliable energies for the LUMO than its ionization potential EOM counterpart for the HOMO. 
The advantage of EA-EOM-pCCD is demonstrated for rylene and rylene diimide units of different chain lengths, where the differences between theoretical and experimental EAs approach chemical accuracy.
\end{abstract}

\maketitle

Over the last few decades, a significant effort has been made to develop alternative technologies to meet the growing demand for electricity. 
One promising example is the design of new organic electronic solar cells with improved power conversion efficiency.~\cite{joule-15-precent-osc, 18-pce-science-2020, opv-19-percent}
Developing new and even more efficient organic devices heavily relies on the fundamental understanding of the charge transfer process between the donor and acceptor units and their mutual alignment.~\cite{cui-osc-review-2020} 
These processes occur at the molecular level, which can be challenging to investigate experimentally. 
To that end, reliable and efficient computational methods are needed to model electronic structures and properties of the building blocks of organic electronic materials.
Among others, particularly important factors are the donor's highest occupied molecular orbital (HOMO) and the acceptor's lowest unoccupied molecular orbital (LUMO) and their offsets. 
The HOMO and LUMO can be directly connected to the ionization potential (IP) and the electron affinity (EA). 
Although those properties can be easily calculated using Density Functional Approximations (DFAs), the results strongly depend on the design of the approximate exchange--correlation functional.~\cite{dft-failure-organic-electronics-acr-2014, dft-failure-tortorella2016,  jahani2023relationship}
This work presents a different approach to determining electron-attached states, resulting LUMO energies, and derived band gaps.
Building on our recent works,~\cite{boguslawski2021open, tecmer2023jpcl} we extend the pair coupled cluster doubles (pCCD) ansatz,~\cite{limacher2013new, stein2014seniority, boguslawski2014efficient, tecmer2022geminal}
\begin{equation}
\ket{\textrm{pCCD}} = e^{\hat{T}_\textrm{pCCD}} \ket{\Phi_0},
\end{equation}
to target open-shell electronic structures through the equation-of-motion (EOM) formalism~\cite{rowe1968equations, stanton1993equation,eom-cc-bartlett2012} for electron attachment,~\cite{nooijen-ea-eom-jcp-1995} where the reference wave function is constrained to electron-pair excitations.
In the above equation, $\big|\Phi_0\rangle$ is some reference determinant and
$\hat{T}_\textrm{pCCD} = \sum_{i}^{n_{\rm occ}} \sum_{a}^{n_{\rm virt}} c_i^a{a_a^\dag}{a_{\bar a}^\dag}{a_{\bar i}}{a_i}$
is the pCCD cluster operator, where $\hat{a}_p$ ($\hat{a}_p^\dag$) are the elementary annihilation (creation) operators for $\alpha$ ($p$) and $\beta$ $(\overline{p})$ electrons, and $c^a_i$ are the pCCD amplitudes.
The sum runs over all occupied $i$ and virtual $a$ orbitals.

%
%
\begin{figure*}[t]
\centering
  \includegraphics[scale=0.6]{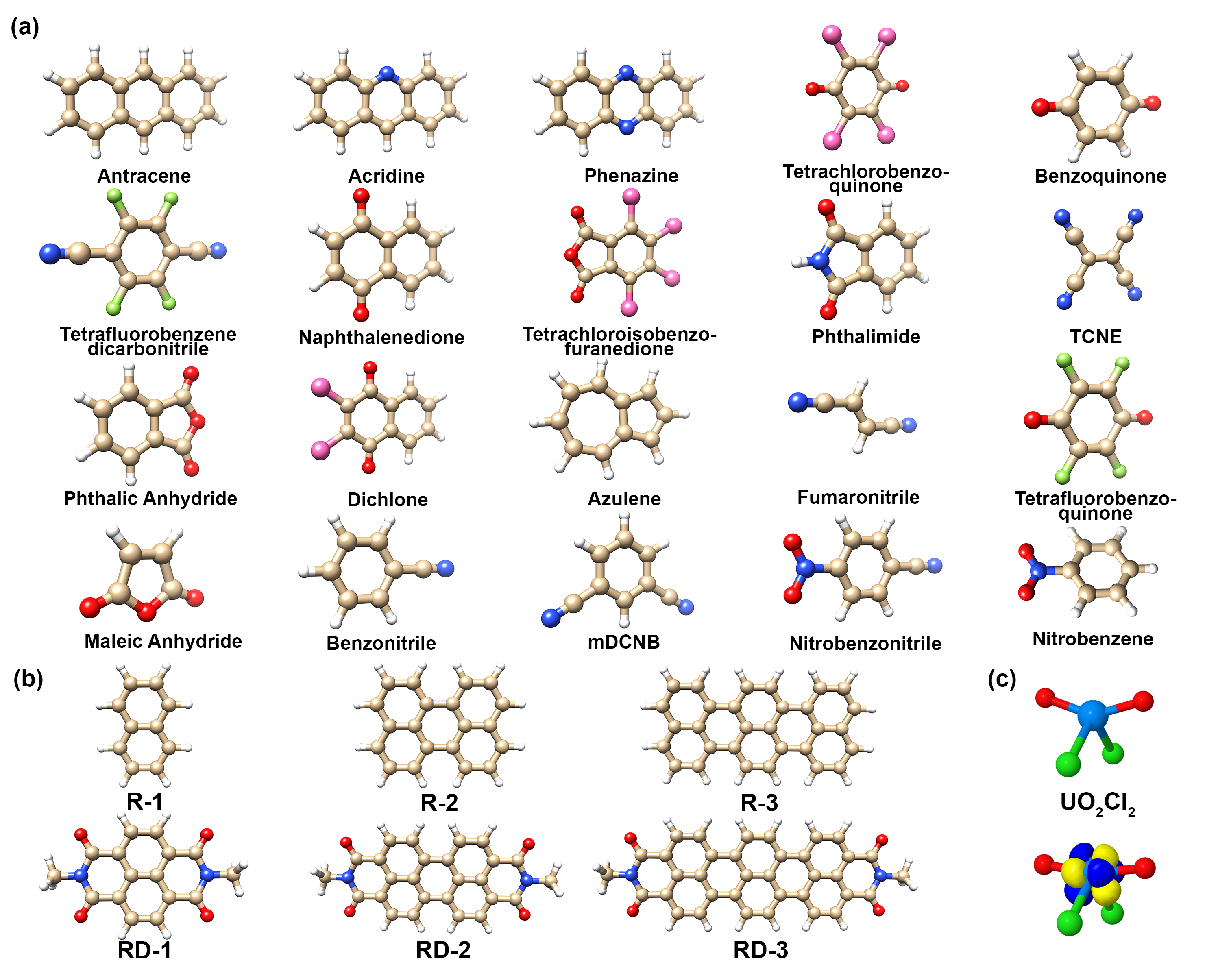}
  \caption{Molecular structures of the three investigated test systems: (a) organic acceptors were relaxed with B3LYP/6-311G** whose molecular geometries are available in the supplementary material of Ref. \citenum{part_three}, (b) the rylene (R) and rylene diimide (RD) molecule series relaxed using B3LYP/cc-pVDZ and (c) UO$_2$Cl$_2$ from Ref.~\citenum{ea-uo2cl2-jms-2021} and the isosurface plot of the unoccupied orbital involed in the electron attachment process of the \ce{^2A1} state of \ce{UO2Cl2-}. See ESI\dag{} for computational details.}
  \label{fig:structures}
\end{figure*}

Starting with the closed-shell pCCD wave function, open-shell electronic structures are created within the EA-EOM formalism by attaching electrons to this closed-shell pCCD reference function through a linear ansatz to parametrize the $k$-th state,\cite{boguslawski2016targeting,boguslawski2017erratum,boguslawski2018targeting,nowak2019assessing,boguslawski2021open}
\begin{equation}
\ket{\Psi_k} = \hat R(k)\ket{\textrm{pCCD}}
\end{equation}
where the operator $\hat R(k)$ generates the targeted state $k$ from the initial pCCD reference state.
The attachment operator $\hat R$ is typically divided into different parts based on the number of particle (electron creation) and hole (electron annihilation) operators contained in each component.
The single EA-EOM formalism~\cite{nooijen-ea-eom-jcp-1995, musial2003equation, musial2014equation} defines $\hat R(k)$ as (dropping the $k$-dependence for reasons of better readability)
\begin{equation}
\hat R^{\rm EA} = \sum_{a}r_a\hat a^\dag_a + \frac{1}{2}\sum_{ab{j}}{r}^{ab}_j\hat{a}_a^\dag \hat a_b^\dag \hat a_j+ \dots ~ = \hat R_{1\rmp}+\hat R_{2\rmp1\rmh}+\dots ~ 
\end{equation}
The attachment states are then obtained by solving the corresponding EOM equations
\begin{equation}
{[\hat{H}_N,\hat R ]} \ket{\textrm{pCCD}} =  \omega \hat R \ket{\textrm{pCCD}},
\end{equation}
where $ \omega = \Delta E - \Delta E_0$ is the energy difference associated with the attachment process with respect to the pCCD ground state,
while $\hat{H}_N = \hat{H} - \langle\Phi_0\big|\hat{H}|\Phi_0\rangle  $ is the normal product form of the Hamiltonian.
The above equation can be rewritten as
\begin{equation}
{\cal H}^\textrm{pCCD}_N \hat R\ket{\Phi_0} =  \omega \hat R \ket{\Phi_0}
\end{equation}
with ${\cal H}^\textrm{pCCD}_N$ being the similarity transformed Hamiltonian of pCCD in its normal-product form ${\cal H}^\textrm{pCCD}_N$= $e^{- \hat{T}_\textrm{pCCD}}\hat{H}_Ne^{\hat{T}_\textrm{pCCD}}$.
The attachment energies are thus the eigenvalues of a non-Hermitian matrix.

Similarly, we can target doubly-electron-attached states through the Double (D)EA-EOM formalism~\cite{gulania2021equation} that extends the $\hat R$ operator to contain at least 2 particle terms,
\begin{align}
\hat R^{\rm DEA}    &= \sum_{ab}r_{ab}\hat a^\dag_a \hat a^\dag_b + \frac{1}{6}\sum_{abc{k}}{r}^{abc}_k\hat{a}_a^\dag \hat a_b^\dag \hat a^\dag_c \hat a_j+ \dots ~ \\ \nonumber
                    &= \hat R_{2\rmp}+\hat R_{3\rmp1\rmh}+\dots ~ 
\end{align}
Here, we assess the performance of two different EA-EOM-based extensions for pCCD.
We focus on the $S_z=0$ and $S_z =-\frac{1}{2}$ (D)EA-EOM flavors ($S_z=0$ for double and $S_z =-\frac{1}{2}$ for single EA).
For the former, the configurational subspace during diagonalization is spanned by
${\ket{\Phi^{a\bb}}, \ket{\Phi^{a\bb c}_{k}}, \ket{\Phi^{a\bb \cb}_{\kb}}}$.
The configurational space for $S_z =-\frac{1}{2}$ simplifies to
${\ket{\Phi^{a}}, \ket{\Phi^{ab}_{j}}, \ket{\Phi^{a\bb}_{\jb}}}$.
The working equations for all EA-EOM and DEA-EOM flavors mentioned above
can be easily derived using diagrammatic techniques, as discussed in
Refs.~\citenum{nooijen-ea-eom-jcp-1995, musial2003equation, musial2014equation}.
All EA-EOM pCCD extensions are implemented in the PyBEST v2.1.0.dev0 software
package.~\cite{boguslawski2021pythonic, boguslawski2024pybest}
{We should stress that such $S_z$ cases can also be efficiently and reliably
treated with spin-flip EOM-CCSD methods.\mbox{\cite{Casanova2020}}}
To illustrate the performance of the proposed EA-EOM-pCCD methods, we investigate three sets of systems shown in Figure~\ref{fig:structures} including (a) a benchmark set of 20 building blocks for organic acceptors, (b) a sequence of rylene (R) and rylene diimide (RD) molecules with an increasing number of naphthalene rings, and (c) the uranyl dichloride molecule.


\begin{table}[t]
\small
  \caption{Statistical error measures [eV], including mean absolute error (MAE) and standard deviation (SD), were assessed based on the IP and EA calculated using IP-EOM-pCCD and EA-EOM-pCCD, respectively. The same analysis was performed for charge gaps ($\Delta_c$) calculated as $\Delta_c = \textrm{IP} - \textrm{EA}$. The experimental data are taken from Refs.~\citenum{Szalay}. The formulas for MAE and SD are printed in the table footnote.
  See ESI\dag{} for computational details.}
  \label{tbl:errors}
  \begin{tabular*}{0.48\textwidth}{@{\extracolsep{\fill}}llll}
    \hline
    & IP & EA & $\Delta_c$ \\
    \hline
    MAE & 1.972 & 0.944 & 2.916\\
    SD & 0.359 & 0.193 & 0.449\\
    \hline
  \end{tabular*}
\begin{tablenotes}
   \item[*] MAE = $\sum_i^N \frac{|E_i^{\rm method} - E_i^{\rm ref}|}{N} $, 
            SD = $\sqrt{\frac{\sum_i^N(E_i^{\rm ME} - \overline{E_i^{\rm ME}})^2}{N}}$
\end{tablenotes}
\end{table}

\begin{figure}[b]
\centering
  \includegraphics[height=3cm]{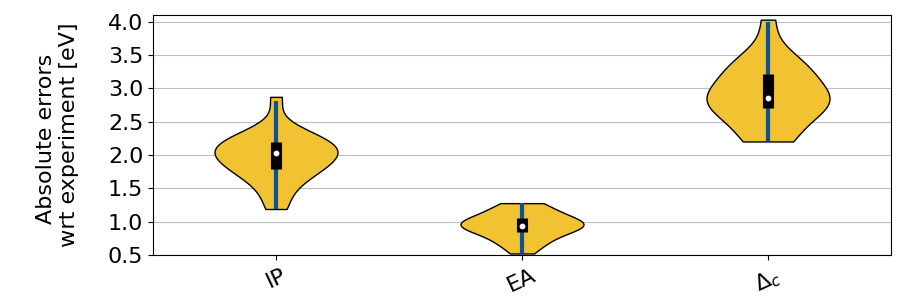}
  \caption{Violin plots illustrating IP- and EA-EOM-pCCD/cc-pVDZ errors w.r.t. the experimental reference set of 20 organic acceptors [eV] shown in Figure~\ref{fig:structures}(a).
  Charge gaps are denoted as $\Delta_c$.
  A white dot in each violin plot represents the median value.
  The IP-EOM-pCCD data is taken from Ref.~\citenum{ip-tailoredpccd}
  See ESI\dag{} for computational details}
  \label{fig:violin-plots}
\end{figure}

Reliable predictions of ionization potentials and electron affinities are crucial for molecules that serve as building blocks for organic photovoltaics. Proper functionalization ensures optimal level alignment between donor and acceptor molecules.
Therefore, our first set of molecules consists of 20 organic acceptors shown in Figure~\ref{fig:structures}(a), including acenes, nitro compounds, nitriles, quinones, and anhydrides.
Ionization potentials for this particular set of molecules were previously investigated using various pCCD-based methods, from the simple IP-EOM-pCCD to various IP-EOM frozen pair CC flavors.\cite{ip-tailoredpccd}
This previous study highlights the importance of dynamical correlation in accurately predicting ionization potentials.
Here, the same set serves as a test case for predicting EA values and the charge gap, $\Delta_c$, calculated as the difference between IP and EA values ($\Delta_c = \text{IP} - \text{EA}$).
A statistical analysis with respect to experimental data,~\cite{Szalay} including mean absolute errors (MAE) and standard deviations (SD) for IP, EA, and $\Delta_c$ calculated within IP- and EA-EOM-pCCD are summarized in Table~\ref{tbl:errors}.
The locality, spread, skewness, and distribution of the errors are shown in the violin plots in Figure~\ref{fig:violin-plots}.
While IP-EOM-pCCD underestimates IPs by 2 eV, the EA-EOM-pCCD counterpart overestimates EAs only by 1 eV with respect to experiment.
However, the errors in IP and EA accumulate, resulting in $\Delta_c$ being larger by approximately 3 eV compared to experimental values.
Nonetheless, the spread of the errors is much smaller for EAs than IPs, where the SD reduces by around 54\%.
Furthermore, the distribution of errors in EAs is strongly centered around the median.
While the IP and $\Delta_c$ errors exhibit relatively symmetric skewness, the EA errors show a slight positive skewness.
Thus, we can conclude that dynamical correlation is less important to predict EA processes within pCCD, while the resulting EA-EOM extension deviates less from the mean (0.2 compared to 0.4 eV).

\begin{table}[t]
\small
    \centering
    \caption{IPs, EAs, and charge gaps ($\Delta_c = \textrm{IP}-\textrm{EA}$) calculated using the IP- and EA-EOM-pCCD methods for the rylene (R) and rylene diimide (RD) series. The differences between the calculated and experimental values ($E(\textrm{pCCD})-E(\textrm{exp})$) are given in parentheses. Experimental results are obtained from Ref.~\citenum{rylene-diimides}. The "$-$" or "$+$" signs indicate that the calculated values are lower or higher than the experimental ones.
    See ESI\dag{} for computational details}
    \label{tbl:rylene}
    \begin{tabular*}{0.48\textwidth}{@{\extracolsep{\fill}}llll}
    \hline
            &  IP & EA & $\Delta_c$ \\
     \hline        
     RD-1    &  6.85($+$0.07)   & 3.11($-$0.52) & 3.74($+$0.59)\\
     RD-2    &  5.35($-$0.66)   & 3.53($-$0.18) & 1.82($-$0.48)\\
     RD-3    &  4.54($-$0.98)   & 3.79($+$0.02) & 0.75($-$1.00) \\
     R-1     &  6.08($+$0.61)   & 0.17($-$1.07) & 5.91($+$1.68) \\
     R-2     &  4.62($-$0.56)   & 1.68($-$0.73) & 2.94($+$0.17) \\
     R-3     &  3.90($-$0.93)   & 2.45($-$0.38) & 1.45($-$0.55) \\
     \hline
    \end{tabular*}
\end{table}

\begin{figure}[b]
\centering
  \includegraphics[height=4cm]{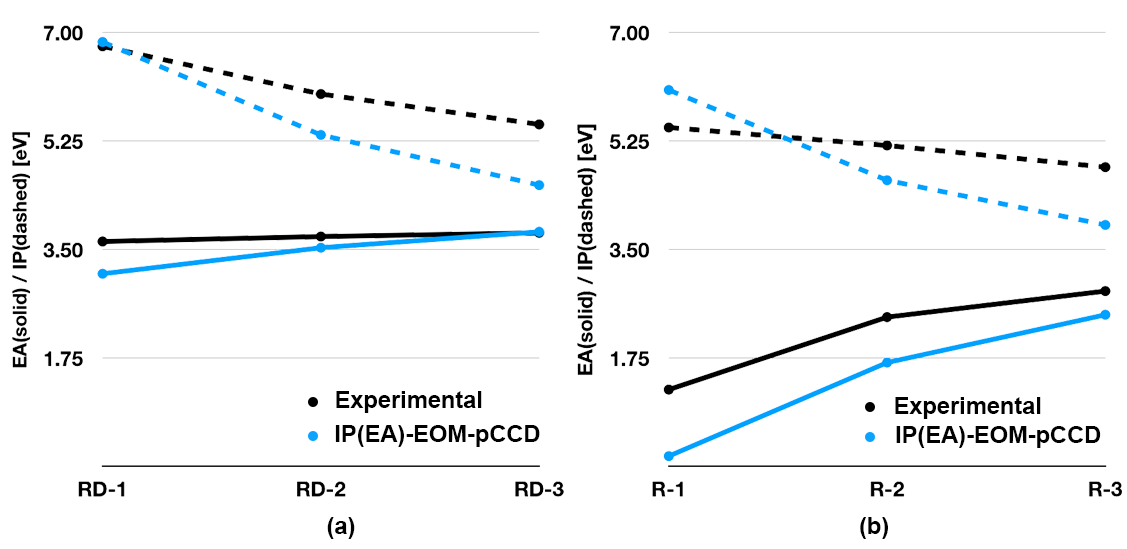}
  \caption{Ionization potentials (dashed lines) and electron affinities (solid lines) calculated with the IP- 
 and EA-EOM-pCCD methods, respectively, for the (a) rylene diimide (RD) and (b) rylene (R) molecules. The theoretical results are marked in blue, while the experimental values are marked in black. The experimental reference data are taken from Ref.~\citenum{rylene-diimides} and included for comparison. They are summarized in the ESI\dag.}
  \label{fig:R_RD}
\end{figure}

Our second test group includes the R and RD family of dyes, shown in Figure~\ref{fig:structures}b. 
Similar to the first group of molecules, they are also excellent acceptor molecules due to their low-lying LUMO levels. 
Their photostability and thermal stability further make them outstanding candidates for various applications in modern electronics. 
The R series is created by connecting additional naphthalene units at their peri positions. 
Thus, R-1 is a simple naphthalene molecule, R-2 has two naphthalene units, R-3 has three, and so on. 
In the RD series, one imide group is attached to each side of the corresponding R molecule (see Figure~\ref{fig:structures}b.) 
The above set of acceptors represents a perfect testing ground to study the systematic dependence of IPs, EAs, and $\Delta_c$ on the number of naphthalene units. 
Table~\ref{tbl:rylene} lists the IP, EA, and $\Delta_c$ values calculated with IP- and EA-EOM-pCCD for the R and RD series with up to three naphthalene units. 
Figure~\ref{fig:R_RD} provides a graphical representation of the evolution of the experimental and theoretical IP and EA for R and RD with increasing ring size. 
Most importantly, EOM-pCCD-based methods reproduce the experimental trend in IP and EA for the R and RD series, respectively.
IPs decrease while EAs increase with the number of naphthalene units and qualitatively resemble the experimental results. 
We should stress, however, that the IPs decrease significantly and approach a different limit than observed in experiment.
The EAs, on the other hand, seem to converge toward the right experimental limit (see Figure~\ref{fig:R_RD}).
This observation demonstrates again the lesser sensitivity of EA-EOM-pCCD to dynamical correlation.
Similarly to the previous benchmark set, the different error scale in IPs and EAs leads to larger errors in charge gaps.


\begin{table}[t]
\small
  \caption{Scalar relativistic adiabatic electron affinities [eV] of the \ce{^2A1} state of \ce{UO2Cl2-} computed from different EOM-pCCD-type methods. 
  The structures are taken from Ref.~\citenum{ea-uo2cl2-jms-2021} and are in the ESI\dag.
  EA-EOM-pCCD exploits the optimized structure of the charged \ce{UO2Cl2-} molecule.
  The electron attachment energy of DEA/EA-EOM-pCCD is computed as the difference in total energies between the DEA-EOM-pCCD ground state of \ce{UO2Cl2} and the corresponding EA-EOM-pCCD \ce{^2A1} state of \ce{UO2Cl2-} to predict the 5f$^1$ (\ce{^2A1}) orbital energy.
  The DIP/IP-EOM-pCCD electron attachment energy is computed as the difference in total energies between the DIP-EOM-pCCD ground state and the IP-EOM-pCCD \ce{^2A1} state of \ce{UO2Cl2^-}. 
  CBS denotes the complete basis set limit. 
  The difference between the reference CCSD(T) value of 2.82 eV and the CBS EOM-pCCD values is given in parenthesis.
  See ESI\dag{} for computational details
  }
  \label{tbl:uranylchloride}
  \begin{tabular*}{0.48\textwidth}{@{\extracolsep{\fill}}llll}
    \hline
    & \multicolumn{3}{c}{\ce{^2A1} \ce{UO2Cl2}} 
    \\ \cline{2-4} 
    & DZ & TZ & CBS  \\
    \hline
    EA-EOM-pCCD     & 5.80 & 6.05 & 6.08($+$3.26)\\
    DEA/EA-EOM-pCCD & 2.32 & 2.27 & 2.25($-$0.57)\\
    DIP/IP-EOM-pCCD & 2.94 & 2.78 & 2.71($-$0.11)\\
    \hline
  \end{tabular*}
\end{table}

Finally, we investigated the first electron attachment of the uranyl dichloride molecule shown in Figure~\ref{fig:structures}c using an EA-EOM-pCCD/ANO-RCC/DKH2 approach.
The uranyl cation, \ce{UO_2^{2+}}, is a common building block of larger uranium-containing molecules.
Its valence electronic structure comprises contributions from atomic uranium 5d, 5f, and 6p orbitals and the oxygen 2p orbitals. 
The uranyl's valence unoccupied (nonbonding) 5f{$_\phi$} and 5f{$_\delta$} orbital sets are occupied when the formal oxidation state of uranium is reduced from +6 to +5 and its electronic configuration changes from 5f$^0$ to 5f$^1$.
The presence of an additional electron on either 5f{$_\phi$} or 5f{$_\delta$} weakens the U--O bond and changes the uranyl-ligand interactions.
To that end, the electron affinity of the neutral uranyl(VI)-based molecules provides valuable information about the 5f orbital energy. 
Table~\ref{tbl:uranylchloride} lists the lowest adiabatic 5f$^1$ (\ce{^2A1}) orbital energy (see Figure~\ref{fig:structures}c) of the uranyl dichloride anion (\ce{UO2Cl2-}) computed with different pCCD-based approaches using structures reported in Ref.~\citenum{ea-uo2cl2-jms-2021}. 
The EA-EOM-pCCD orbital energy is far from the CCSD(T) reference value of 2.82 eV~\cite{ea-uo2cl2-jms-2021} and is most likely related to the missing dynamical correlation effects in the model.
The DEA/EA-EOM-pCCD and DIP/IP-EOM-pCCD alternatives minimize these effects (as we compare two different EA or IP flavors with each other) and, thus, provide more reliable orbital energies, where the DIP/IP error in orbital energies approaches approximately 2.5 kcal/mol. 

To sum up, our numerical examples highlight that the presented EOM-pCCD methods yield much smaller errors for EAs than IPs.
As demonstrated for the R and RD series, the theoretically predicted EAs gradually reach the experimental values with an increasing number of naphthalene rings, approaching chemical accuracy.
On the other hand, the atomic character of the \ce{UO2Cl2} lowest attached state poses a greater challenge to EA-EOM-pCCD, which can be alleviated by switching to a DEA/EA-EOM-pCCD and DIP/IP-EOM-pCCD description. 
Further improvements in the accuracy of the pCCD-based EA models can be achieved by incorporating dynamical correlation effects via, for example, frozen-pair pCCD variants.


\section*{Data availability}
The data supporting this article have been included as part of the ESI\dag.

\section*{Conflicts of interest}
There are no conflicts to declare.

\section*{Acknowledgements}
M.G.~acknowledges financial support from a Ulam NAWA -- Seal of Excellence research grant (no.~BPN/SEL/2021/1/00005). 
P.T.~acknowledge financial support from the SONATA BIS research grant from the National Science Centre, Poland (Grant No. 2021/42/E/ST4/00302). 
We acknowledge that the results of this research have been achieved using the DECI resource Bem (Grant No.~412) based in Poland at Wroclaw Centre for Networking and Supercomputing (WCSS, http://wcss.pl) with support from the PRACE aisbl. 
Funded/Co-funded by the European Union (ERC, DRESSED-pCCD, 101077420).
Views and opinions expressed are, however, those of the author(s) only and do not necessarily reflect those of the European Union or the European Research Council. Neither the European Union nor the granting authority can be held responsible for them.

\bibliography{rsc}
\bibliographystyle{rsc} 

\end{document}